\documentclass[%
 aip,
 jcp,%
 amsmath, amssymb,
 reprint,%
]{revtex4-1}

\usepackage{xcolor}

\usepackage{hyperref}   
\hypersetup{
	colorlinks = True,
	allcolors = {blue}
}
\usepackage{graphicx}
\usepackage{dcolumn}
\usepackage{breqn}

\makeatletter
\let\cat@comma@active\@empty
\makeatother

\begin{document}


\title{On the difference between variational and unitary coupled cluster theories}

\author{Gaurav Harsha}
\affiliation{Department of Physics and Astronomy, Rice University, Houston, Texas 77005, USA}

\author{Toru Shiozaki}
\affiliation{Department of Chemistry, Northwestern University, Evanston, Illinois 60208, USA}

\author{Gustavo E. Scuseria}
\affiliation{Department of Chemistry, Rice University, Houston, Texas 77005, USA}
\affiliation{Department of Physics and Astronomy, Rice University, Houston, Texas 77005, USA}


\begin{abstract}
There have been assertions in the literature that the variational and unitary forms of coupled cluster theory lead to the same energy functional. Numerical evidence from previous authors was inconsistent with this claim, yet the small energy differences found between the two methods and the relatively large number of variational parameters precluded an unequivocal conclusion. Using the Lipkin Hamiltonian, we here present conclusive numerical evidence that the two theories yield different energies. The ambiguities arising from the size of the cluster parameter space are absent in the Lipkin model, particularly when truncating to double excitations. We show that in the symmetry adapted basis under strong correlation the differences between the variational and unitary models are large, whereas they yield quite similar energies in the weakly correlated regime previously explored. We also provide a qualitative argument rationalizing why these two models cannot be the same. Additionally, we study a generalized non-unitary and non-hermitian variant that contains excitation, de-excitation and mixed operators with different amplitudes and show that it works best when compared to the traditional, variational, unitary, and extended forms of coupled cluster doubles theories.
%
\end{abstract}


\maketitle

\section{\label{sec:intro}Introduction}
Coupled cluster (CC) theory\cite{crawford,RevModPhys.79.291} has established itself as a benchmark model in computational quantum chemistry. It has been used to compute fairly accurate approximations to the energy eigenvalues of many body systems with reasonable computational cost.
The ground state energy in the traditional CC (tCC) framework is given by
\begin{equation}
    E_{\text{CC}} = \left \langle 0 \right |~ e^{-T} H e^{T} ~ \left | 0 \right \rangle,
    \label{cc-energy}
\end{equation}
where $T$ is the excitation cluster operator truncated to $k^{\text{th}}$ rank ($T = T_1 + T_2 + ... + T_k$). Different values of $k$ yields different truncated CC \emph{Ansatze}.
Traditional CC theory comes with the perks of size extensivity through the exponential \emph{Ansatz}, as well as fairly good approximations to the ground state energy of a wide range of Hamiltonians, particularly when weakly correlated. However, being an asymmetric expectation value, it is not an upper bound to the energy.

On the other hand, the variational CC (vCC)\cite{Szalay1995,Voorhis2000} energy is obtained by minimizing the symmetric expectation value
\begin{equation}
    E_{\text{vCC}} = \frac{\left \langle 0 \right |~ e^{T^\dagger} H e^{T} ~\left | 0 \right \rangle}{\left \langle 0 \right |~ e^{T^\dagger} e^{T} ~\left | 0 \right \rangle}
    \label{vCC}
\end{equation}
and is a rigorous upper bound. Like traditional CC, vCC is size extensive in the sense that it contains only contributions from connected diagrams.\cite{Pal1982,Pal1983}
However, it yields an infinite series of contributing terms with no systematic truncation scheme. This makes vCC computationally unpractical for realistic systems.
An alternative coupled cluster approach based on a unitary \emph{Ansatz},\cite{Kutzelnigg1991,BARTLETT1989133,taube2006new} known as unitary CC (uCC) has been sought as a bridge between traditional CC and vCC. The uCC energy is obtained by minimizing the expectation value
\begin{equation}
    E_{\text{uCC}} = \left \langle 0 \right |~ e^{-\sigma } H e^{\sigma } ~ \left | 0 \right \rangle,
    \label{uCC}
\end{equation}
where $\sigma = T-T^\dagger$ is an anti-hermitian operator. Unitary CC is, by definition, a symmetric expectation value. Analogous to vCC, an infinite series of diagrams contribute to $E_{\text{uCC}}$. Recently, there has been a renaissance of interest in uCC theory in connection with quantum simulators that can evolve unitary operators.\cite{Romero2017}

It has been asserted in the literature\cite{Pal1982} that the energy functional for the variational and unitary forms of coupled cluster theory result in identical expressions- in fact identical order by order in the expansion. On the other hand, ref [\onlinecite{Kutzelnigg1991}] reports otherwise. Numerical studies\cite{cooper2010,evangelista2011alternative} that followed this claim reported discrepancies between $E_{\text{vCC}}$ and $E_{\text{uCC}}$. Ref [\onlinecite{cooper2010}] reported that for the molecule Hydrogen Flouride, at the separation length $R=3\mathrm{\AA}$ the vCC and uCC energies differ by 2 milliHartree or more. Yet the small differences in the two energies, as well as the relatively large number of parameters in the variational optimization (which makes it tedious and usually difficult to converge to the true global minimum), has precluded an unequivocal resolution about the validity of the claim. A proper analysis and comparison of the two energy functional is thus warranted.

In this paper, we make use of the Lipkin\cite{LIPKIN1965188, MESHKOV1965199, GLICK1965211} model hamiltonian to present concluding numerical evidence that the variational ($E_{\text{vCC}}$) and the unitary ($E_{\text{uCC}}$) coupled cluster energies are different.
Moreover, we offer a qualitative yet concrete argument justifying the observed discrepancies.
We also compare different variants of coupled cluster doubles (CCD) model - uCCD, vCCD, tCCD, extended\cite{Arponen1987,PIECUCH1999295,FAN20061} CCD (eCCD), and a generalized\cite{Nooijen2000,NooijenPRL,Nakatsuji2000,FAN20061} variant (gCCD) containing independent excitation, de-excitation and mixed operators that we solve variationally (vgCCD).

This article has been structured in the following way: in Sec. \ref{sec:lip}, we describe the Lipkin model, highlighting the relevant features that enable numerical tests to be performed conveniently.
Sec. \ref{sec:meth} lays out the details of the numerical schemes utilized while in Sec. \ref{sec:res} we present the results obtained using these schemes at the doubles level- comparing energies obtained with vCCD, uCCD, tCCD, eCCD, and vgCCD. In Sec. \ref{sec:qual}, we present a qualitative argument justifying the difference observed between variational and unitary coupled cluster theories. We close with concluding remarks.

\section{\label{sec:lip}The Lipkin Model}
\begin{figure}[b]
	\centering
    \includegraphics[width=0.75\linewidth]{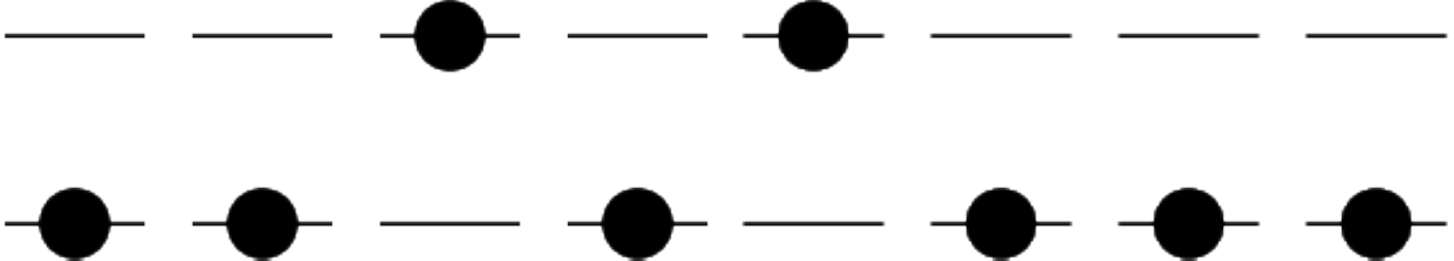}\\
    \vspace{0.1in}
    {\small (a) Even parity.}\\
    \vspace{0.1in}
    \includegraphics[width=0.75\linewidth]{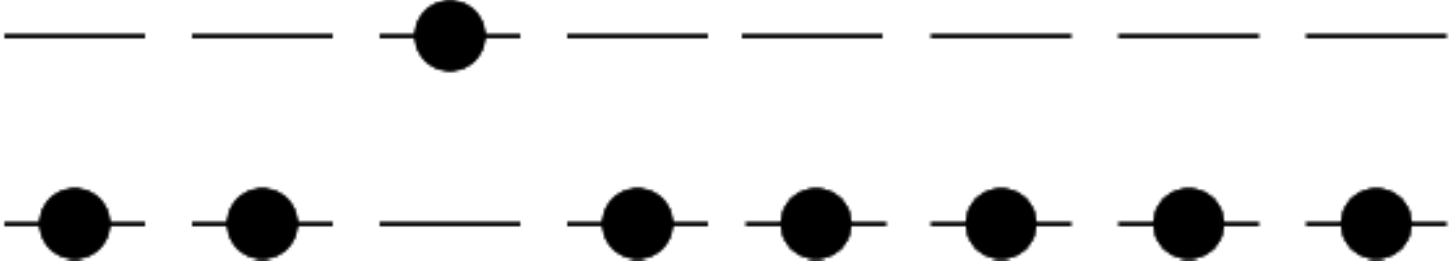}\\
    \vspace{0.1in}
    {\small (b) Odd parity.}
    \caption{Typcial configurations with even and odd parity for N=8 site Lipkin model.}
    \label{fig:lipkin}
\end{figure}

The Hamiltonian due to Lipkin, Meshkov, and Glick is a simple yet non-trivial model describing fermions in a two level system. It was originally proposed as a model to describe a closed shell nucleus with schematic monopole interactions.\cite{LIPKIN1965188, MESHKOV1965199, GLICK1965211}
There are $N$ sites at each of the two levels and exactly $N$ fermions that are allowed to hop between levels (excitations and de-excitations) but not between different sites. Fig.~\ref{fig:lipkin} shows two of the many possible configurations for a $N=8$ site Lipkin model.
The Hamiltonian exhibits \emph{su(2)} symmetry and can be mathematically expressed in terms of \emph{su(2)} generators
\begin{equation}
    H = x J_z - \frac{1-x}{N} \left ( J_+^2 + J_-^2 \right ).
\end{equation}
For quantum chemistry purposes, it is convenient to write the generators in terms of spin 1/2 creation and annihilation fermion operators $J_z$, $J_\pm$ defined as
\begin{align}
    J_z &= \frac{1}{2}\sum_{i=1}^{N}~ c_{i\uparrow}^\dagger c_{i\uparrow} - c_{i\downarrow}^\dagger c_{i\downarrow}, \\
    J_+ &= \sum_{i=1}^{N}~ c_{i\uparrow}^\dagger c_{i\downarrow}, \\
    J_- &= J_+^\dagger,
\end{align}
which satisfy \emph{su(2)} commutation relations
\begin{equation}
    \left [J_+,J_-\right ] = 2J_z \quad \text{and} \quad \left [J_z,J_\pm \right ] = \pm J_\pm.
\end{equation}
Here $x$ conveniently parameterizes the relative strength of the interaction: $x = 1$ is non-interacting while $x \rightarrow 0$ is the strong correlation limit. 
It is not difficult to see that $J^2$ is a symmetry and hence we can label the eigenstates of the system by quantum number $j$ with integral values of $j$ such that $j(j+2)/4$ are the eigenvalues of $J^2$. For the $N$ site Lipkin model, any configuration or state lies in $j=N$ subspace and this further simplifies our problem. Another symmetry of the Hamiltonian is parity defined as $\hat{P} = e^{i\pi J_z}$. As an example, for $N=8$, the states with even number of excited fermions have the parity eigenvalue $p = 1$ (see Fig.~\ref{fig:lipkin}a) and those with odd number of fermions in the upper level have $p = -1$ (see Fig.~\ref{fig:lipkin}b).

The Lipkin model is exactly solvable using the Richardson-Gaudin \emph{Ansatz}\cite{LERMAH2013421,ORTIZ2005421} as well as using the generator coordinate method.\cite{ring2004nuclear} However, exact diagonalization (full configuration interaction) of the resulting banded Hamiltonian is straightforward for fairly large system sizes and is the approach used herein. This is a consequence of the size of the Hilbert space being merely $N+1$ as opposed to the combinatorial dependence of the size on number of particles in typical many body systems such as the Hubbard model. We conveniently form an orthonormal basis based on the number of fermions excited to the upper level and define
\begin{equation}
|m\rangle = J_+^m |0\rangle
\end{equation}
$\forall m = 0,1,2,...,N$. It must be noted that while there are $^NC_m$ possible states with $m$ fermions excited, all of them are degenerate and therefore, $|m\rangle$ is, by definition, a linear combination of all these configuration each with a coefficient $(^NC_m)^{-1/2}$ when normalized. Despite the simplicity, the model contains non-trivial physics in the transition between the weakly and strongly correlated regimes where parity symmetry breaks down at the Hartree-Fock level.\cite{jacob2017}

The restricted Hartree-Fock (RHF) reference determinant for the Lipkin Hamiltonian is given by
\begin{equation}
\left | 0 \right \rangle  = \prod_i c_{i\downarrow}^\dagger \left | - \right \rangle,
\end{equation}
where $|-\rangle$ is the physical vacuum with no particles. One can also form a broken symmetry determinant given by
\begin{equation}
\left | \Phi \right \rangle = \prod_i \alpha_{i\downarrow}^\dagger \left | - \right \rangle,
\end{equation}
where $\{\alpha_i\}$ defines the deformed basis of creation or annihilation operators, obtained through a unitary rotation of $\{c_i\}$
\begin{equation}
\left ( \begin{matrix}
\alpha_{i\uparrow}^\dagger \\
\alpha_{i\downarrow}^\dagger
\end{matrix} \right ) = \frac{1}{\sqrt{1 + \kappa^2}} \left ( \begin{matrix}
1 & \kappa \\
-\kappa & 1
\end{matrix} \right ) \left ( \begin{matrix}
c_{i\uparrow}^\dagger \\
c_{i\downarrow}^\dagger
\end{matrix} \right ).
\label{uhf-eq}
\end{equation}
Such a transformation with a non-trivial value for $\kappa$ breaks the parity symmetry of the Hamiltonian.
The above equation can be conveniently re-written as a Thouless rotation\cite{THOULESS1960225}
\begin{equation}
\left | \Phi \right \rangle = \frac{1}{(1 + \kappa^2)^{N/2}} e^{\kappa J_+} \left | 0 \right \rangle
\end{equation}
where the value of $\kappa$ is optimized so as to minimize the Hartree-Fock broken symmetry energy, here loosely referred to as unrestricted HF (UHF), in analogy with quantum chemistry terminology for broken spin symmetry
\begin{equation}
E_{\mathrm{UHF}} = \left \langle \Phi \right | H \left | \Phi \right \rangle.
\end{equation}
In our study, we make use of both UHF and RHF references to study the difference between vCC and uCC. However, as we shall observe later, this difference is more clearly apparent in the RHF.

Requiring that the excitation operator respects the parity symmetry of the Hamiltonian,\cite{jacob2017} the cluster excitation operator is defined as:
\begin{equation}
    T = \sum_{i=1}^{k}~ t_i J_+^i.
\end{equation}
Note that there is only one amplitude per excitation level due to the degeneracy. These features make the Lipkin model an ideal tool for comparing approximate many body theories. \cite{jacob2017,Mazziotti2004,ARPONEN1983141,EMRICH1981397,Robinson1989}
Both the variational as well as the traditional coupled cluster approaches are fairly easy to converge.

\section{\label{sec:meth}Methods}
In this study, we make use of different coupled cluster \emph{Ansatze}: traditional, variational, unitary, extended, and generalized. The latter is a non-hermitian coupled cluster \emph{Ansatz} similar to generalized CC where we use different amplitudes for excitation, de-excitation, and mixed combination operators; we obtain the energy variationally as an expectation value. Here, we shall describe the explicit mathematical formulation only for the RHF reference; it is straightforward to generalize the equations for a UHF reference. In the results, we refer to the coupled cluster methods based on the RHF and a UHF reference as RCC and UCC respectively. Accordingly, traditional CCD, for instance, becomes tRCCD and tUCCD.


\subsection{Traditional and extended coupled cluster}
Practical CC models will necessitate some level of truncation. We therefore here restrict the cluster operator $T$ to double excitations.
While working with the RHF reference, we use an \emph{Ansatz} that preserves the symmetry of the wave function and thus, the cluster operator consists of double excitations only: $T = tJ_+^2$. On the other hand, for UHF reference, we use CCD including additional single excitations (CCSD). The traditional coupled cluster doubles energy (tCCD) in the RHF  basis is given by
\begin{align}
    E_{\mathrm{tCCD}} &= \left \langle 0 \right |~e^{-tJ_+^2} H e^{tJ_+^2} ~ \left | 0 \right \rangle, \\
    &= \left \langle 0 \right |~ \bar{H}~\left | 0 \right \rangle , 
    \label{CCD-energy}
\end{align}
where $\bar{H}$ is the similarity transformed Hamiltonian and the doubles amplitude $t$ for tCCD  is found by solving
\begin{equation}
    0 = \left \langle 0 \right |~J_-^2~ \bar{H} ~ \left | 0 \right \rangle.
    \label{CCD-ampl-eq}
\end{equation}
The extended coupled cluster energy\cite{Arponen1987,PIECUCH1999295,FAN20061} is obtained by a second similarity transformation of $\bar{H}$, this time with de-excitation operators
\begin{equation}
    E_{\mathrm{eCCD}} = \left \langle 0 \right | ~ e^{zJ_-^2} e^{-tJ_+^2} H e^{tJ_+^2} e^{-zJ_-^2} ~\left | 0 \right \rangle.
    \label{eCC-energy}
\end{equation}
After noting that de-excitations to the right do not change the reference, it is also possible to interpret eCC as tCC with a more elaborate bra state. Note that $\bar{H}$ is non-hermitian, thus its right and left eigenvectors must be different
\begin{equation}
E_{\mathrm{eCC}} = \left \langle 0 \right |~e^{Z} \bar{H} ~\left | 0 \right \rangle.
\label{eCC-energy2}
\end{equation}
Improvements to the bra yield significant changes to tCC. The amplitudes for eCCD are obtained by making the energy functional in eq.~(\ref{eCC-energy2}) stationary with respect to all parameters, i.e. $t$ and $z$.

For a standard Hamiltonian like Lipkin, with at most 2 body terms, the Baker-Campbell Hausdorff expansion of equation (\ref{CCD-energy}) truncates at the fourth order nested commutator yielding a non-hermitian 6-body effective Hamiltonian. For eCC, equation (\ref{eCC-energy}) yields an even higher many-body effective H but still truncates at $T_2^4$ and $Z_2^3$.

\subsection{Variational and unitary \emph{Ansatze}}

The variational coupled cluster doubles (vCCD) energy is obtained by minimizing the symmetric (hermitian) expectation value
\begin{equation}
    E_{\mathrm{vCCD}} = \underset{t}{\mathrm{min}} \frac{\left \langle 0 \right |~e^{tJ_-^2} H e^{tJ_+^2} ~ \left | 0 \right \rangle}{\left \langle 0 \right |~ e^{tJ_-^2} e^{tJ_+^2}~ \left | 0 \right \rangle}.
    \label{vCCD}
\end{equation}
The variational approach to CC is, in general, computationally intractable because it involves a number of terms that grow combinatorially as a function of size. In the Lipkin model, however, it is possible to converge the vCC wave function robustly because of the relatively small size of the Hilbert space. In our numerical implementation, we evaluate the vCCD wave function using a Taylor series expansion of the exponential of the double excitation operator:
\begin{equation}
    \left | \mathrm{vCCD} \right \rangle = \left ( 1 + tJ_+^2 + \frac{t^2}{2\!}J_+^4 + \frac{t^3}{3\!}J_+^6 + ... \right )\left | 0 \right \rangle.
\end{equation}
The Taylor series is truncated to $k^{\mathrm{th}}$-order when the norm contribution from the $(k+1)^{\mathrm{th}}$ order is within a specified tolerance value. For a model with $N$ sites, the Taylor series naturally truncates at $N^{\mathrm{th}}$ power of $J_+$ since
$$ J_+^{N+1} | 0 \rangle = 0.$$
The hermitian expectation value of the Hamiltonian matrix can then be easily obtained with order $N$ operations. The resulting functional is minimized with respect to $t$ in a one-dimensional optimization.

Analogously, the uCCD energy is obtained by taking the symmetric expectation value of $H$ using the exponential of an anti-hermitian operator of double excitation and de-excitation operators, $\sigma = t~(J_+^2 - J_-^2)$, where $\sigma^\dagger = -\sigma$
\begin{equation}
    E_{\mathrm{uCCD}} = \underset{t}{\mathrm{min}} \left \langle 0 \right |~e^{-\sigma} H e^{\sigma} ~ \left | 0 \right \rangle.
    \label{uCCD}
\end{equation}
Computationally, unitary coupled cluster can be implemented using an approach similar to the variational \emph{Ansatz}. However, one can easily diagonalize the anti-hermitian operator $\sigma' = J_+^2 - J_-^2$, which has imaginary eigenvalues
$$
    \sigma' \left |\lambda_p \right \rangle = i \lambda_p \left | \lambda_p \right \rangle,
$$
where we demand the eigenvalues $\lambda_p \in \mathbf{R}$ by explicitly inserting $i$; writing the RHF reference in the basis of $\{ | \lambda_p \rangle \}$
\begin{equation}
    |0\rangle = \sum_p c_p |\lambda_p\rangle ,
\end{equation}
the energy expression then takes the form
\begin{equation}
    E_{\mathrm{uCCD}} = \underset{t}{\mathrm{min}} \sum_{p,q} c_p^* c_q e^{-i (\lambda_p - \lambda_q)} \left \langle \lambda_p \right |~H~\left | \lambda_q \right \rangle.
\end{equation}
For both the variational and unitary CCD energy functionals, a well defined global minimum with respect to $t$ is observed. It must be noted that the variational energy functional in the UHF picture will have an additional dependence on $\kappa$ (see equation (\ref{uhf-eq})) and one should also minimize the energy with respect to this parameter.

\subsection{Generalized \emph{Ansatz}}
Our last model is a variational \emph{Ansatz} in the spirit of generalized coupled cluster, where we make use of $J_+^2$, $J_-^2$ and $J_+J_-$ operators. In principle, one can also include $J_z$. But we observe that including the latter causes a strong tendency for the amplitudes to converge to values such that
\begin{equation}
\exp (tJ_+^2 + aJ_-^2 + bJ_z) \rightarrow \exp (-\beta H)
\end{equation}
with large values for $\beta$ ($>1$) leading to numerical issues. This observation is similar to the one made by Mazziotti in one of the schemes in Ref [\onlinecite{Mazziotti2004}].
The energy for this variational-generalized coupled cluster doubles (vgCCD) is given by
\begin{align}
    \left | \mathrm{vgCCD} \right \rangle &= \exp (tJ_+^2 + aJ_-^2 + bJ_+J_-) \left | 0 \right \rangle, \\
    E_{\mathrm{vgCCD}} &= \underset{t,a,b}{\mathrm{min}} \frac{\left \langle \mathrm{vgCCD} \right |~H ~ \left | \mathrm{vgCCD} \right \rangle}{\left \langle \mathrm{vgCCD} | \mathrm{vgCCD} \right \rangle}.
\end{align}
Computationally, the vgCCD energy in the symmetry adapted basis is obtained using an approach similar to that explained above for vCCD. However, it is difficult to find a global minimum in the broken symmetry formulation.
Even with RHF as the reference, care must be taken to get the required convergence. Here we follow a systematic cascade approach- restricting the amplitude $b=0$, we conveniently minimize the energy with respect to $t$ and $a$; these converged values are then used as an initial guess and energy is minimized with respect to all the three parameters $\{t,a,b\}$.

\begin{figure}[t]
    \centering
    \includegraphics[width=\linewidth]{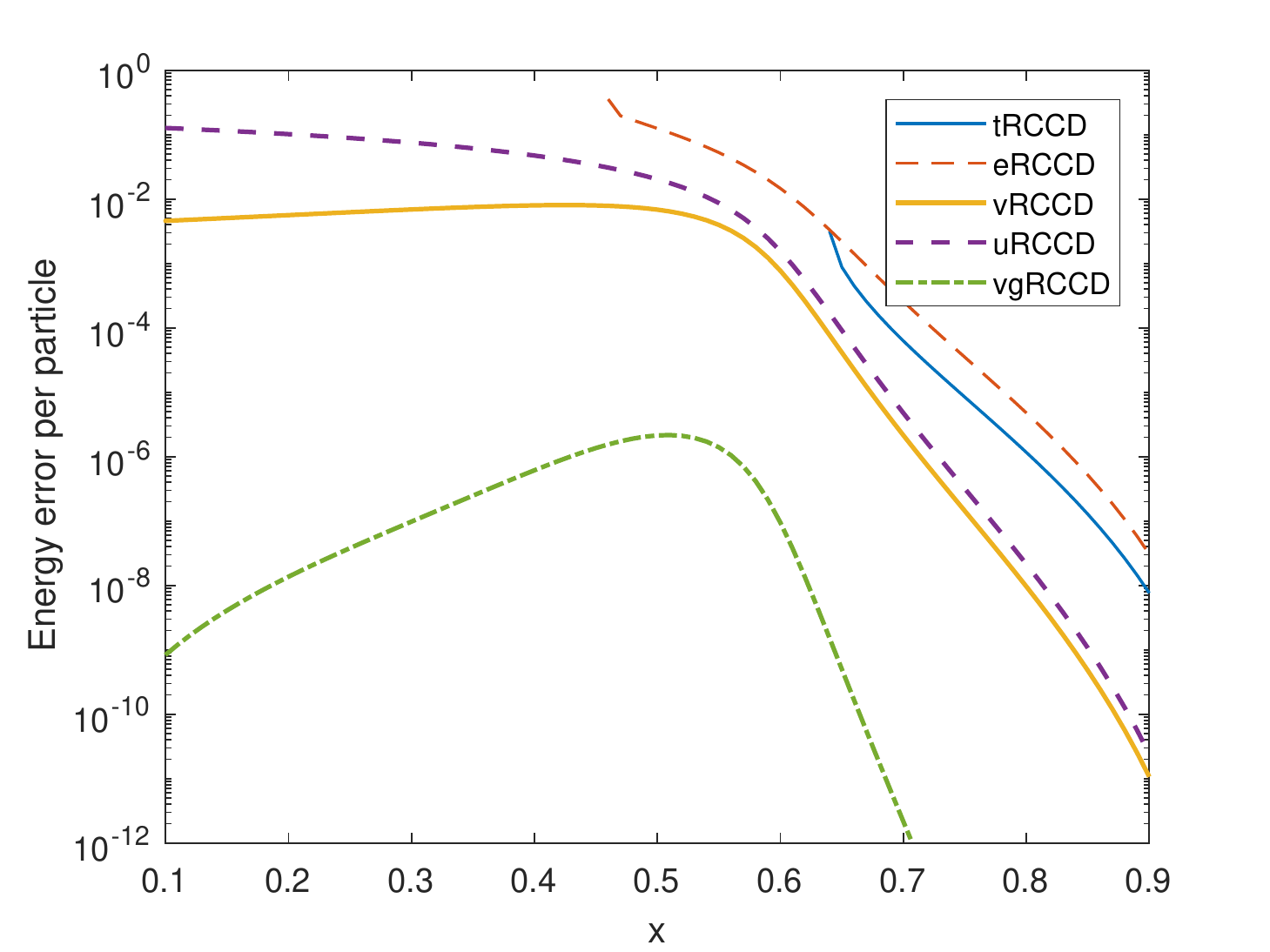}
    \caption{Energy error per particle for traditional, extended, variational, unitary, and variational-generalized CCD in the RHF basis on Lipkin model for $N=32$ sites.}
    \label{fig:rccd}
\end{figure}

\section{\label{sec:res}Results}
The various \emph{Ansatze} described above were implemented for the Lipkin model without any significant numerical problems. Energy plots as a function of the interaction parameter $x$ are presented for both symmetry adapted as well as broken symmetry formulations. Keeping consistent with notation in Ref [\onlinecite{jacob2017}], we use a prefix `R' for the results in the RHF basis results and a prefix of `U' for those in the UHF basis. Figure~\ref{fig:rccd} shows energy error per particle with respect to exact results for the various methods in the RHF basis i.e., traditional (tRCCD), extended (eRCCD), variational (vRCCD), unitary (uRCCD) and generalized (vgRCCD) coupled cluster doubles for $N=32$ site Lipkin model. As already noted in previous work,\cite{jacob2017} both the tRCCD and eRCCD solutions fail to exist as the correlation strength increases ($x$ decreases).

\begin{figure}[t]
    \centering
    \includegraphics[width=\linewidth]{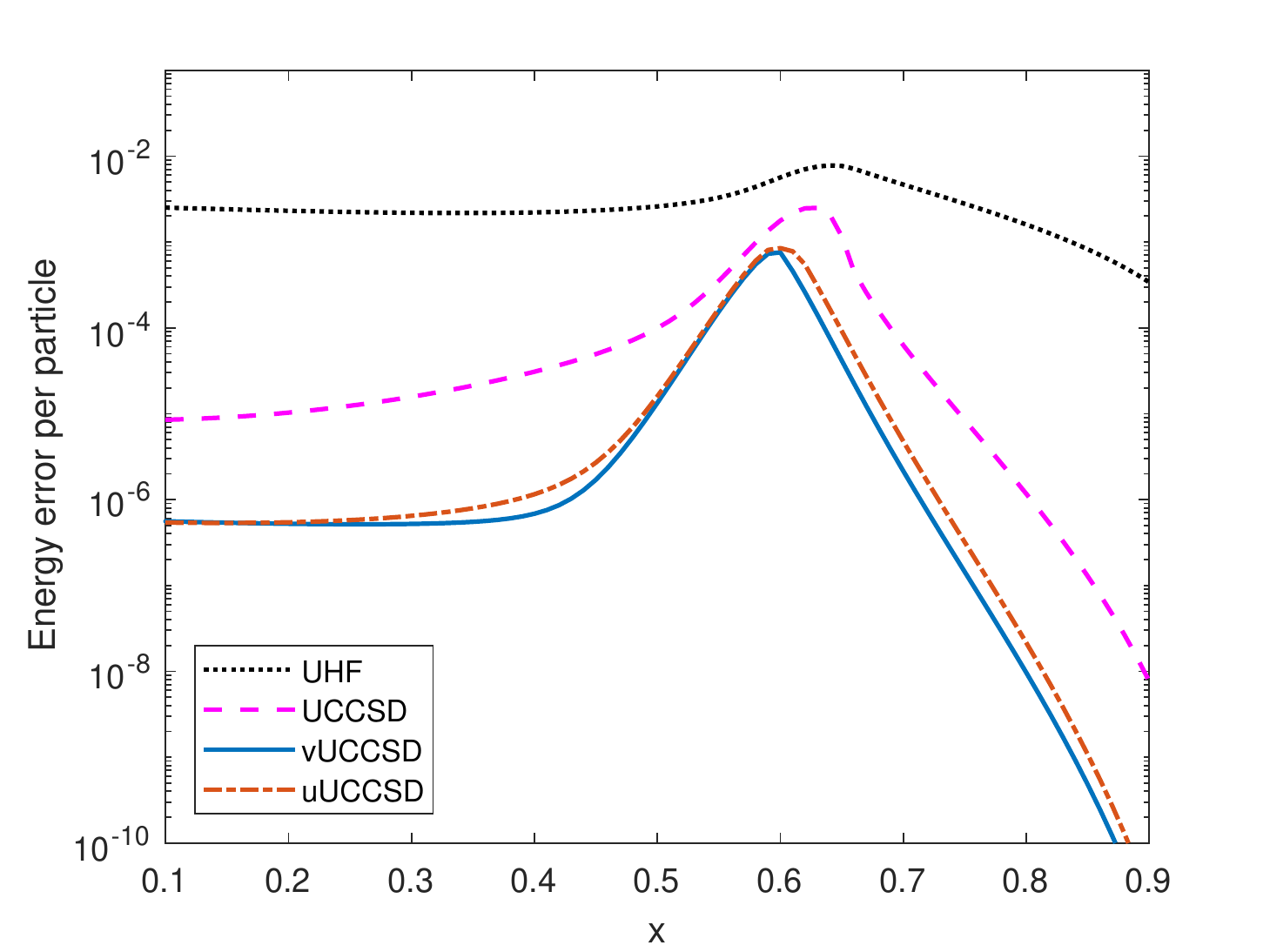}
    \caption{Energy error per particle for mean field energy, traditional, variational and unitary CCSD in the UHF basis on Lipkin model for $N=32$ sites. The variational energy functionals are optimized with respect to $\kappa$ as well.}
    \label{fig:uccsd}
\end{figure}

\begin{figure}[t]
    \centering
    \includegraphics[width=\linewidth]{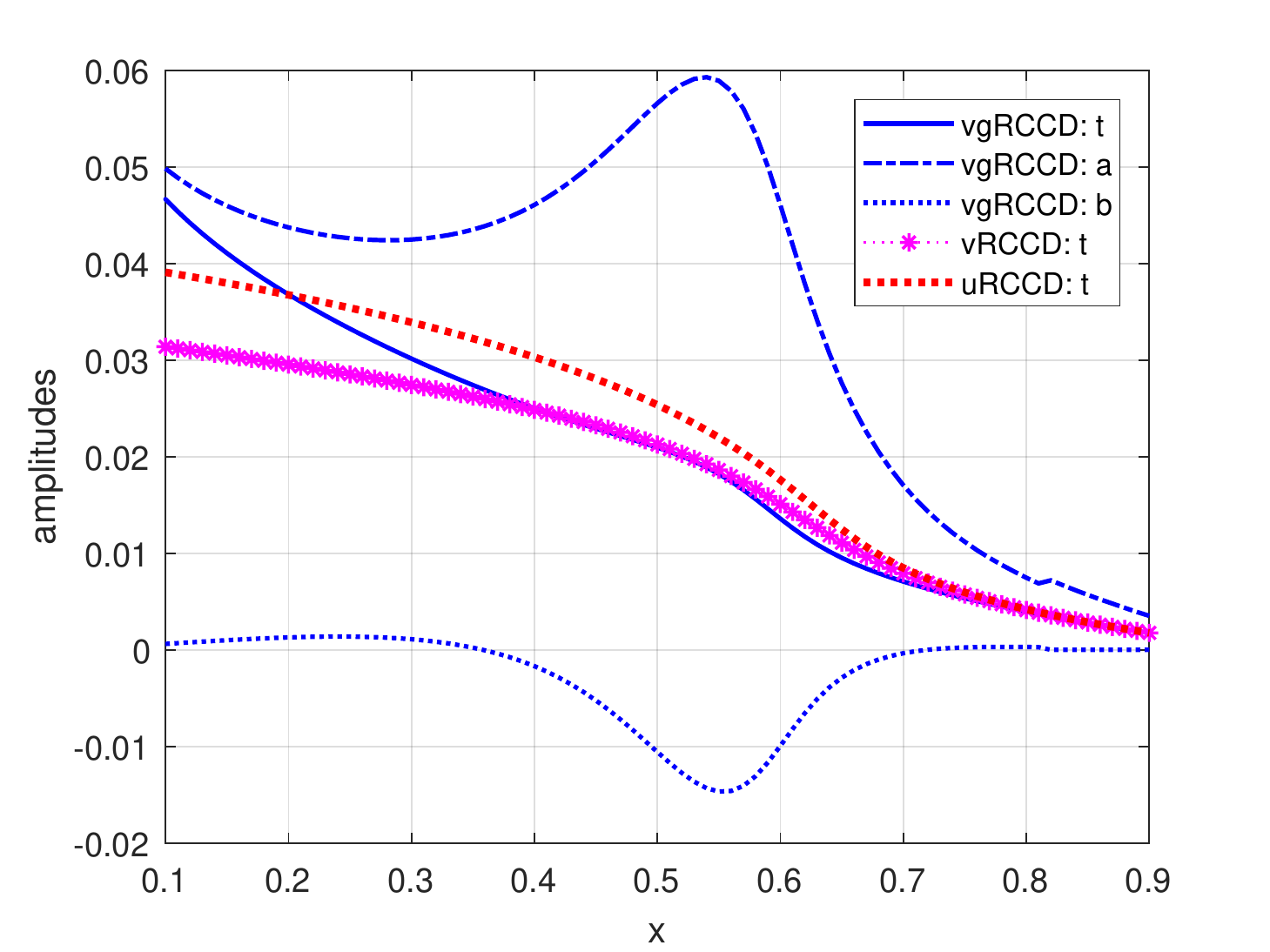}
    \caption{Optimized cluster amplitudes corresponding to the different vRCCD methods in Fig~\ref{fig:rccd} for $N = 32$ Lipkin model.}
    \label{fig:gvamps}
\end{figure}

Figure~\ref{fig:uccsd} shows energy error per particle with respect to exact results for the various methods in the broken symmetry basis i.e., UHF energy, traditional (tUCCSD), variational (vUCCSD) and unitary (uUCCSD) coupled cluster doubles for $N=32$ site Lipkin model. As noted before, for the variational methods, we minimize the energy functional with respect to the broken symmetry parameter $\kappa$ as well.
Figure~\ref{fig:gvamps} shows the optimized amplitudes corresponding to the different RCCD \emph{Ansatze} in Fig~\ref{fig:rccd}. Clearly, unitary and variational \emph{Ansatze} do not lead to the same energy - in fact, vRCCD is found to be better than uRCCD over the entire range of the interaction parameter $x$ for reasons that are not obvious to us. The difference is not as visible in the broken symmetry formulation. We give a qualitative explanation for this observation in the Sec. \ref{sec:qual}. On the other hand, the vgRCCD \emph{Ansatz} significantly improves upon both vRCCD and uRCCD. Moreover, the optimum amplitudes for vgRCCD make the cluster operator non-hermitian and far from unitary.

\section{\label{sec:qual}Variational vs unitary}
It is not difficult to see the difference between the vCC and uCC energy functionals analytically.
As noted by Pal \emph{et. al.} in Ref [\onlinecite{Pal1982}], the vCC energy functional can be reduced to
\begin{equation}
    E_{\mathrm{vCC}} = \sum_{m,n} \frac{1}{m!n!} \langle 0 |~ (T^\dagger)^m H T^n ~|0 \rangle_{\mathrm{cl}},
    \label{vCC-func}
\end{equation}
where the subscript `cl' denotes contribution from closed and connected terms (or diagrams) only. On the other hand, the uCC energy functional can be expanded using the Baker-Campbell-Hausdorff series
\begin{equation}
    E_{\mathrm{uCC}} = \sum_{k} \frac{1}{k!} \langle 0 |~ [[...[H,\sigma],\sigma],...],\sigma] ~|0 \rangle,
    \label{uCC-func}
\end{equation}
where the summand is the ground state expectation of the usual $n^{\mathrm{th}}$ order nested commutator of $H$ and the unitary cluster operator $\sigma = T-T^\dagger$. At $k^{\mathrm{th}}$ order nested commutator of $H$ with $\sigma$, we have $^kC_m$ unique combinations of nested commutators of $H$ with a string of $m$ $T$'s and $(k-m)$ $T^\dagger$'s $\forall m = 1,2,...,n$. If these commutators are either all trivial or all non-trivial, then we recover the coefficients in eq.~(\ref{vCC-func}) with $n = k -m$.

In general, however, the BCH expansion of an arbitrary term $h_i$ in the Hamiltonian with either pure excitation or pure de-excitation operators will terminate at a finite order, say, $n_{i+}$ and $n_{i-}$ respectively. Then at some $(m < n_{i\pm})^{\mathrm{th}}$ order in the eq.~(\ref{uCC-func}), the ground state expectation value of a nested commutator of $h_i$ with some arbitrary sequence of $x$ $T$'s and $y$ $T^\dagger$'s ($x+y = m$) becomes
\begin{multline}
\frac{1}{m!} \left \langle 0 \right |~[...[[H,T],T^\dagger],...T^\dagger]~\left | 0 \right \rangle = \\
\frac{1}{m!} \left \langle 0 \right |~(T^\dagger)^y H T^x \left | 0 \right \rangle_{\mathrm{cl}}
\end{multline}
and since $m$ can be partitioned as $m=x+y$ in $m!/(x!y!)$ ways, the contribution from $h_i$ at $m^{\mathrm{th}}$ order to uCC in eq.~(\ref{uCC-func}) and vCC in eq.~(\ref{vCC-func}) is identical, as has been claimed. However, for $m>n_i$ some of the $m!/(x!y!)$ combinations would be identically zero, which result into the difference between these two energy functionals.

As an explicit example, consider the term $h = J_+^2$ in the Lipkin Hamiltonian and the corresponding contribution at $O(t^3)$ in vCCD and uCCD. The only closed connected term in vCCD is
\begin{equation}
E_{\mathrm{vCCD}}^{(3)}(h) = \frac{t^3}{1!2!} \left \langle 0 \right | (T^\dagger)^2 h T \left |0 \right \rangle_{\mathrm{cl}},
\label{vcc_3rd}
\end{equation}
while the uCCD functional due to $h$ at the same order is
\begin{multline}
E_{uCCD}^{(3)}(h) = \frac{t^3}{3!}  \Big\{ [[[h,T],T^\dagger],T^\dagger] + [[[H,T^\dagger],T^\dagger],T]  \\
+[[[H,T^\dagger],T],T^\dagger]  \Big\}.
\end{multline}
Notice that there are $3!/(1!2!)$ terms in the above expression for uCCD, but clearly the first term is zero while the others are equivalent to the closed and connected expectation value in eq.~(\ref{vcc_3rd}). Therefore
\begin{equation}
E_{vCCD}^{(4)} = \frac{3}{2}E_{uCCD}^{(4)}
\label{diff3_ucc_vcc}
\end{equation}
Another example that shows the difference at $O(t^4)$ can also be constructed with $h_z=J_z$. While $\left \langle 0 \right | (T^\dagger)^2 h_z T^2 \left |0 \right \rangle$ contributes non-trivially to vCCD, the following commutators in the uCC expression are zero
\begin{equation}
[[[[h_z,T],T,T^\dagger],T^\dagger] = 0 =
[[[[h_z,T^\dagger],T^\dagger],T],T].
\end{equation}
Indeed, for the Lipkin model Hamiltonian, the variational and unitary energy functional would differ by $O(t^3)$. Therefore, if the optimizing amplitudes $t_{\mathrm{vCC}}$ and $t_{\mathrm{uCC}}$ are similar (and here found to be $\sim10^{-2}$ - see Fig~\ref{fig:gvamps}), then the observed difference in $E_{\mathrm{uCC}}$ and $E_{\mathrm{vCC}}$ would be $\sim10^{-6}$ or lower. This is exactly what we observe in the broken symmetry formulation. On the other hand, in the symmetry adapted Hamiltonian, the optimizing amplitudes $t_{\mathrm{vCC}}$ and $t_{\mathrm{uCC}}$ are quite different. Thus we see a more clear difference between the two energy functionals.

\section{\label{sec:conc}Conclusions}
We have unequivocally demonstrated the non-equivalence of energies between unitary and variational CCD \emph{Ansatze} using the Lipkin model Hamiltonian. The difference between uCC and vCC is found to be small in weakly correlated limit. This is consistent with observations made in Refs [\onlinecite{cooper2010},\onlinecite{evangelista2011alternative}] where physical systems have been studied in weak correlation limit and the energy differences are reported to be of $O(mE_h)$ or even smaller. It is in strongly correlated limit when the distinction between the two energy functionals is evident (see Fig.~\ref{fig:rccd}). This observation is corroborated by our arguments in Sec. \ref{sec:qual} where we notice that for the interaction Hamiltonian, the energy difference shows up at $O(t^3)$ while for non-interacting case, it shows up at $O(t^4)$. Furthermore, the results from the variational-generalized \emph{Ansatz} suggests that a non-hermitian combination of $J_+^2$, $J_-^2$ and $J_+J_-$ operators leads to much smaller errors than the anti-hermitian one. While we have explicitly worked with the Lipkin Hamiltonian, it is straightforward to generalize the arguments presented in Sec. \ref{sec:qual} to the more general \emph{ab initio} Hamiltonian.

\begin{acknowledgements}
This work was supported by the US Department of Energy, Office of Basic Energy Sciences, Computational and
Theoretical Chemistry Program under Award No. DE-FG02-09ER16053. G.E.S. acknowledges support as a Welch Foundation
Chair (C-0036). We thank Peter Knowles and Francesco Evangelista for helpful discussions.
\end{acknowledgements}

\bibliography{newbib2}

\end{document}